# Optical Vortex Ladder via Sisyphus Pumping of Pseudospin

Sihong Lei[1,*], Shiqi Xia[1,*], Daohong Song[1,2], Jingjun Xu[1], Hrvoje Buljan[1,3], Zhigang Chen[1,2]

1 The MOE Key Laboratory of Weak-Light Nonlinear Photonics, TEDA Institute of Applied Physics and School of Physics,

Nankai University, Tianjin 300457, China

2 Collaborative Innovation Center of Extreme Optics, Shanxi University, Taiyuan, Shanxi 030006, China

3 Department of Physics, Faculty of Science, University of Zagreb, Bijenička c. 32, Zagreb 10000, Croatia

*These authors contributed equally to this work.

songdaohong@nankai.edu.cn, hbuljan@phy.hr, zgchen@nankai.edu.cn

**Robust higher-order optical vortices are much in demand for applications in optical manipulation, optical communications, quantum entanglement and quantum computing. However, in numerous experimental settings, a controlled generation of optical vortices with arbitrary orbital angular momentum (OAM) remains a substantial challenge. Here, we present a concept of "optical vortex ladder" for stepwise generation of optical vortices through Sisyphus pumping of pseudospin modes in photonic graphene. Instead of conical diffraction and incomplete pseudospin conversion under traditional Gaussian beam excitations, the vortices produced in the ladder arise from non-trivial topology and feature diffraction-free Bessel profiles, thanks to the refined excitation of the ring spectrum around the Dirac cones. By employing a periodic "kick" to the photonic graphene, effectively inducing the Sisyphus pumping, the ladder enables tunable generation of optical vortices of any order even when the initial excitation does not involve any OAM. The optical vortex ladder stands out as an intriguing non-Hermitian dynamical system, and, among other possibilities, opens up a pathway for applications of topological singularities in beam shaping and wavefront engineering.**

Optical vortices, featuring wavefront screw dislocations or phase singularities, have captivated great attention over the decades, particularly in connection with the orbital angular momentum (OAM) of light[1-3]. A host of fascinating phenomena have been demonstrated with optical vortex beams, from unusual diffraction patterns[4] to toroidal vortices of light[5], and from twisted photons[6,7] to spin-orbit interactions[8-10]. Vortex generation has become a flourishing research area, driven by enormous applications including optical tweezers[11-13], quantum information processing[14-16], optical communications[17,18] and vortex lasers[19-21]. Despite technological advancements in optical vortex generation such as spiral phase plates[22,23], liquid crystal devices[24,25], metasurfaces and nanophotonics[19,26,27], the complex structural design and susceptibility to fabrication defects set avoidable constraints and limitations in many OAM-based applications.

Over the past decade, topological photonics has opened a promising avenue for creating robust modes that are immune to impurities, disorders and defects[28,29]. The study of topological materials has sparked interdisciplinary interest[30-34], leading to discoveries of new photonic topological edge modes[35-37], higher-order topological phases[38,39] and disclination states[40-42], as well as Weyl and Dirac semimetals[43-45]. Photonic systems with momentum-space topological singularities have also been utilized for vortex generation through the spin-orbit interaction[46,47]. Due to the topological nature of the Dirac singularities, the vortex generation is immune to disorder and impurities in those systems and does not require precise alignment of the incident beam relative to the structures[47], unlike in other cases with phase plates and metasurfaces. As such, topological systems are particularly attractive for the generation of optical vortices on demand. However, realization of higher-order vortices requires many of Berry phase windings around topological singularities in momentum space, which are typically limited by the symmetry and design of microstructures [46,47].

Here, we propose and demonstrate a scheme to construct an optical vortex ladder (OVL). In successive steps of propagation through periodically "kicked" photonic graphene, a probe beam transforms into vortices with the topological charge increasing one by one in every step of the ladder (Fig. 1). The key ingredients of this process arise from a ring-spectrum excitation around the Dirac cone, involving (i) a complete conversion of pseudospin to OAM in every step of the ladder (where the pseudospin corresponds to sublattice degree of freedom), and (ii) Sisyphus pumping of the pseudospin achieved by a judicious "kick" applied on the photonic graphene at specific propagation lengths. This results in diffraction-free propagation of vortices during the conversion process, akin to the Bessel beams, which meanwhile benefit momentum-space topological protection arising from the Berry phase winding around topological singularities. Our experimental results of vortex generation to any order with preserved OAM through the sequential ascension of the ladder stages are corroborated by theoretical analysis and are broadly applicable to other Dirac-like systems.

**Theoretical analysis for a complete pseudospin conversion**

We describe the pseudospin-orbit interaction in photonic graphene (inset in Fig. 1b), which consists of two sublattices ($A$ and $B$) in one unit-cell. The propagation of the light is governed by the Schrödinger-type equation in the paraxial approximation. Furthermore, under the tight-binding approximation, the equation becomes the two-band Hamiltonian

$$H(\mathbf{k}) = \begin{bmatrix} 0 & t(1 + e^{i\mathbf{k}\mathbf{a}_2} + e^{i\mathbf{k}\mathbf{a}_1}) \\ t(1 + e^{-i\mathbf{k}\mathbf{a}_2} + e^{-i\mathbf{k}\mathbf{a}_1}) & 0 \end{bmatrix}, \quad (1)$$

where $t$ is the nearest neighbor coupling, $\mathbf{a}_1, \mathbf{a}_2$ are the two basic vectors shown in Fig. 1b, and $\mathbf{k} = (k_x, k_y)$. Photonic graphene features a conical intersection at two inequivalent Dirac points (**K** and **K′**) and the topological features around these points are clearly visible in the effective Hamiltonian:

$$H = \frac{\kappa}{2}(\sigma_x p_x + \sigma_y p_y), \quad (2)$$

where $\sigma_i$ represents the components of the pseudospin angular momentum operator, $p_x$ and $p_y$ indicate the displacements of the transverse wavevectors with respect to the Dirac point **K**, $\kappa = \sqrt{3}at$ is related to the coupling coefficient $t$ and lattice constant $a$. The eigenmodes around the Dirac cone are $\psi_{n,\mathbf{k}} = (\phi_A, \phi_B)^T = (n, \exp(i\theta_k))^T/\sqrt{2}$, where $\phi_A$, $\phi_B$ represent the components on the two sublattices, $k_x + ik_y = |\mathbf{k}| \exp(i\theta_k)$ and $n = \pm 1$ denotes the band number. The Berry phase winding number around the Dirac point is $w = \pi$, which can be viewed as a momentum-space topological singularity[46].

In graphene, pseudospin corresponds to the sublattice degree of freedom: pseudospin up (down) corresponds to the $A$ ($B$) sublattice[10]. The eigenmodes of the pseudospin operator $\sigma_z/2$ are $\chi_\uparrow = (1,0)^T$ and $\chi_\downarrow = (0,1)^T$, which relate to the Hamiltonian eigenmodes: $\chi_\uparrow = \psi_{+,\mathbf{k}} - \psi_{-,\mathbf{k}}$, and $\chi_\downarrow \exp(i\theta_k) = \psi_{+,\mathbf{k}} + \psi_{-,\mathbf{k}}$. When we excite a single pseudospin component (say, in sublattice $A$), during propagation, a vortex is generated in the other sublattice ($B$) due to the mapping of topological singularity from momentum to real space[46]. In the previous works, the Gaussian-type probe beams were commonly used for pseudospin mode excitations[9,46]. However, for such excitation methods, the probe beams occupy all Bloch modes around the Dirac point and fail to achieve the complete conversion required by the OVL. Therefore, in this work, we redesign the probing conditions and, for the first time, employ a judicious excitation with a ring spectrum around the Dirac cone. We initially excite all modes with eigenvalues $\Delta\beta/2$ (ring at the upper cone) and $-\Delta\beta/2$ (ring at the lower cone, see Fig. 2a1) and if such an excitation occurs solely on the $A$ sublattice (pseudospin up), then it is straightforward to show that the wavefunction evolution can be written as $\Phi(z) = \sum_{\mathbf{k}}[\chi_\uparrow \cos(\Delta\beta z/2)\exp(i\mathbf{kr}) + i\chi_\downarrow \sin(\Delta\beta z/2)\exp(i(\mathbf{kr} + \theta_k))]$, where $\Phi(z)$ excites both top and bottom bands around the Dirac cone, see Supplementary Section 1.

This type of dynamics serves as the basis for constructing an OVL, in which a complete conversion of pseudospin up to pseudospin down (and vice versa) occurs at specific propagation distances set by the eigenvalue difference $\Delta\beta$. The amount of power residing in pseudospin up is $\gamma_\uparrow(z) = \langle\Phi(z)|P_A|\Phi(z)\rangle \propto \cos^2(\Delta\beta z/2)$, whereas that in pseudospin down is $\gamma_\downarrow(z) = \langle\Phi(z)|P_B|\Phi(z)\rangle \propto \sin^2(\Delta\beta z/2)$, as illustrated in Fig. 2b1. Here, $P_{A(B)}$ is the projection operator on $A$ ($B$) sublattice, i.e., the real-space spin-up (spin-down) projector (Supplementary Section 1). This serves as the first key ingredient for the OVL. The ring-spectrum excitations manifest as Bessel beam excitations in real space (Fig. 1b). The length $L = \pi/\Delta\beta$ of each stage in the OVL can be flexibly adjusted by tuning the diameters of the ring spectra, i.e., by tuning $\Delta\beta$ (Supplementary Section 3). The pseudospin conversion is topologically protected due to the nontrivial Berry phase winding around the Dirac cone[46].

The second ingredient for the OVL is Sisyphus pumping. When the probe beam is exactly at the end of stage N, all the power is in the pseudospin-down mode. Then, all that power is pumped into the pseudospin-up mode (Fig. 2a2), initiating the Sisyphus pumping process. In the tight-binding approximation, the pumping is equivalent to multiplying $\Phi(z)$ by the ladder operator $\begin{pmatrix} 0 & 1 \\ 0 & 0 \end{pmatrix}$ in the OVL, which corresponds to the pseudospin raising operator and raises up the eigenvalue of pseudospin components in $\Phi(z)$. Hence, the power residing in pseudospin up mode, which gradually increases throughout the stages, suddenly drops to zero at the end of the stage (blue line in Fig. 2b2), reminiscent of the Sisyphus process sketched in Fig. 1a. Since the topological charge carried by the pseudospin up modes in stage $N + 1$ is inherited from that of the pseudospin down modes in stage $N$ (Fig. 2a2), the topological charge of the probe beam increases by one at every stage (Fig. 2b2). The corresponding calculations in Fig. 2c demonstrate the topological charge carried by the probe beam at the end of each stage, further confirming the feasibility of the OVL generation.

**Experimental demonstration of periodic pseudospin conversion**

To establish a photonic graphene lattice, we employ the multi-beam optical induction method[48] (see more details in Supplementary Section 4). A collimated laser beam (with a power of 100mW and wavelength of 532nm) is split into three broad beams (quasi-planewaves), which points towards three **K** points in momentum space and form a triangle intensity pattern in real space. Two sets of triangle patterns are sent in turn to a nonlinear crystal (with a refractive index of $n_0 = 2.35$) for inducing the two sublattices of photonic graphene. An electric field ($200 kVm^{-1}$) applied across the crystal along the crystalline *c*-axis leads to the photorefractive effect[48], which combines and translates the two triangle patterns into a coupled honeycomb waveguide array (photonic graphene). The nearest spacing between waveguides is approximately $9.6 \mu m$, as depicted in Fig. 3a. The probe beam is constructed with an overall Bessel distribution using a spatial light modulator (SLM) for excitation of the three **K** points in the first Brillouin zone of the lattice. Since all components of the probe beam carry the same topological information[46], we only need to extract the interferogram from one of the components at the lattice output.

When the probe beam ($l = 0$) with diameter $d_1 = 0.16|\mathbf{K}|$ in momentum space (Fig. 3b3) is launched to cover sublattice $A$, it excites the pseudospin-up mode (zeroth-order Bessel envelope Fig. 3b1). After a 20mm-long propagation through the lattice, the output exhibits a first-order Bessel envelope with topological charge $l' = 1$, indicating a complete pseudospin conversion (Fig. 3b2). By tuning the diameters of the ring spectra, the probe beam is constructed with a larger ring in momentum space $d_2 = 0.32|\mathbf{K}|$ (Fig. 3c4) and smaller lobes in real space (Fig. 3c1). Now the propagation distance required to achieve a complete pseudospin conversion under this probe condition is exactly half of that for the $d_1 = d_2/2$ beam. To verify this, we change to use a 10mm-long crystal and capture the first-order Bessel envelope ($l' = 1$) at the back facet of the crystal (Fig. 3c2). To show the periodicity, we use the same experimental parameters as those for the 20mm-long experiment. The profile of the output beams returns to the initial mode distribution with $l' = 0$ after 20mm-long propagation (Fig. 3c3). Corresponding simulations and direct comparisons with Gaussian beam excitations are presented in the Supplementary Sections 2 and 4, which agree well with the experimental results and theoretical analysis. Thus, we demonstrate clearly the periodic pseudospin conversion under the Bessel beam excitations. These findings provide essential ingredients in the OVL construction.

**Scheme for Sisyphus pumping in photonic graphene**

To realize the Sisyphus pumping in experiment, we employ a periodically "kicked" structure, as shown in Fig. 4. If a pseudospin-up mode with a Bessel envelope is excited at sublattice $A$, as illustrated in Fig. 4a, it distributes completely on sublattice $B$ after completing the pseudospin conversion in stage I. Before entering stage II, the lattice is shifted so that sublattice $A$ maps exactly onto sublattice $B$ of the previous stage, which is fully equivalent to pumping pseudospin down to pseudospin up. The lattice shifting follows the rule:

$$V_n = V_0(\mathbf{r} - (n-1)(\mathbf{a_1} + \mathbf{a_2})/3), \qquad (3)$$

where $n$ is the number of steps in the OVL and $V_0(\mathbf{r})$ represents the refractive-index potential of photonic graphene, which is the same as that in stage I (Fig. 1a). It is evident that $H(z) = H(z + 3L)$, where $H(z)$ is the time-dependent Hamiltonian, the system returns to its initial position after $3L$, i.e., $V_{IV} = V_I$. Such a sudden shifting ("kicking") of the structures allows for the Sisyphus pumping and continuous mapping of topological singularity from momentum to real space[48], which in turn leads to the OVL formation with output optical vortices of any order. Moreover, even when perturbations are applied on the nearest-neighbor couplings, the pseudospin conversion preserves (Supplementary Section 5), and the Bessel beam excitations enable self-healing property of the generated vortices against obstructions. These features give rise to the robustness of the OVLs (Supplementary Section 2).

It should be noted that periodic "kicking" of the graphene lattice ($V_n \rightarrow V_{n+1}$) implies that the dynamics in this Hamiltonian should generally be non-Hermitian as the power is not conserved. For our specific initial conditions and with a proper design of eigenvalue difference $\Delta\beta$ according to the length of the crystal, there is no loss of power as pseudospin-raising operator always acts on the system when it is in the lower pseudospin mode. Moreover, it is worth highlighting that the OVL possesses topological protection owing to the nontrivial winding around the Dirac point. Even when perturbations are introduced in the couplings that affect the system, causing a portion of energy to be trapped in the initial sublattices, the robustness of nontrivial winding and pseudospin conversion persists. This trapped energy is scattered into the higher bands due to non-Hermitian properties introduced by the Sisyphus pumping between the stages. These scattered modes can be interpreted as losses around the Dirac cones but will not impede the pseudospin conversion in the subsequent stage (Supplementary Section 5). Therefore, in spite of the mode sensitivity often encountered in non-Hermitian systems[49], the combined effects of topological and non-Hermitian properties of periodically "kicked" photonic graphene contribute to the robust pseudospin conversion in the OVL.

**Experimental realization of the OVL**

In the experiment, the effectiveness of the OVL is limited by the length of the crystal and the lattice "kicking" between different stages. To overcome this limitation, we utilize a cascade-probing technique. Specifically, the amplitude and phase of the output beam from one waveguide section are captured using a camera, digitally replicated with a SLM, and then sent back into the next section of waveguide arrays. By transferring all amplitude and phase information from one section to another, this cascade-probing technique effectively connects all parts of the waveguide arrays, allowing us to observe the mode evolution through the otherwise length-limited lattice structure.

To initiate the OVL, we generate a probe beam with a zeroth-order Bessel envelope ($l = 0$) using an SLM. This beam excites the pseudospin-up mode at the beginning of stage I (Fig. 4a1). At the end of stage I ($z = 20\,\text{mm}$), the probe beam fulfills the complete pseudospin conversion ($l' = 1$), transferring most energy to sublattice $B$ (Fig. 4a2). In stage II, we apply the cascade-probing method to duplicate the output beam from stage I, resending it to cover sublattice $A$ but aiming at the three **K** points in momentum space where the phase singularity ($l = 1$) is inherited (Fig. 4b1). This process achieves the Sisyphus pumping, so the output of stage II has a topological charge $l' = 2$ (Fig. 4b2). In stage III, the probe beam duplicated from the output of stage II carries the charge $l = 2$ located on sublattice $A$ (Fig. 4c1). After achieving the complete pseudospin conversion, the topological charge increases to $l' = 3$ (Fig. 4c2). In the subsequent stages, the "kicked" lattice enters the next cycle, and the topological charge carried by the probe beam continues to increase. In principle, optical vortices of any order can be obtained through the OVL. These experimental results confirm the effectiveness of the OVL, despite the impurity of the lattice structures (Fig. 3a) and the slight mismatch due to the cascade-probing method (Fig. 4).

## Conclusion

In summary, we have proposed and demonstrated the generation of an OVL that can produce vortices with an arbitrary OAM. This is achieved by topological singularity mapping around the Dirac cones with ring-spectrum excitation and periodic "kicking" of the graphene lattice, which leads to Sisyphus pumping of the pseudospin modes. The concept of the "kicking" effect may also be utilized for successive incremental focusing[50], offering a new approach for future studies in light manipulation. In contrast to electro-optical elements arranged in series in real space, which require alignment and are sensitive to defects[51], the vortex generation in the OVL shows robust properties, featuring diffraction-free Bessel profiles and momentum-space topological protection. The Sisyphus pumping process, along with relevant non-Hermitian properties associated with the OVL, heralds a promising technique for OAM generation and light-field manipulation. We believe the exploration of alternative setups based on this approach could unlock the full potential for practical use in generating high-order OAM in all-optical compact devices. In particular, the OVLs may open exciting new avenues for application of topological singularities, with the potential to impact the fields of optical communication[17,21,52], quantum information computing[14,53] and complex structured light control[54,55]. Our results pave the way for further exploration of topological systems for active and tunable generation and manipulation of optical vortices, as well as for the development of innovative technologies in the field of photonics and beyond.

**Acknowledgments**:

This work was supported by National Key R&D Program of China (No. 2022YFA1404800); the National Nature Science Foundation of China (No. 12134006, 12274242, and 12204252); the Natural Science Foundation of Tianjin (No. 21JCJQJC00050) and the 111 Project (No. B23045) in China. H.B. acknowledges support by the QuantiXLie Center of Excellence.

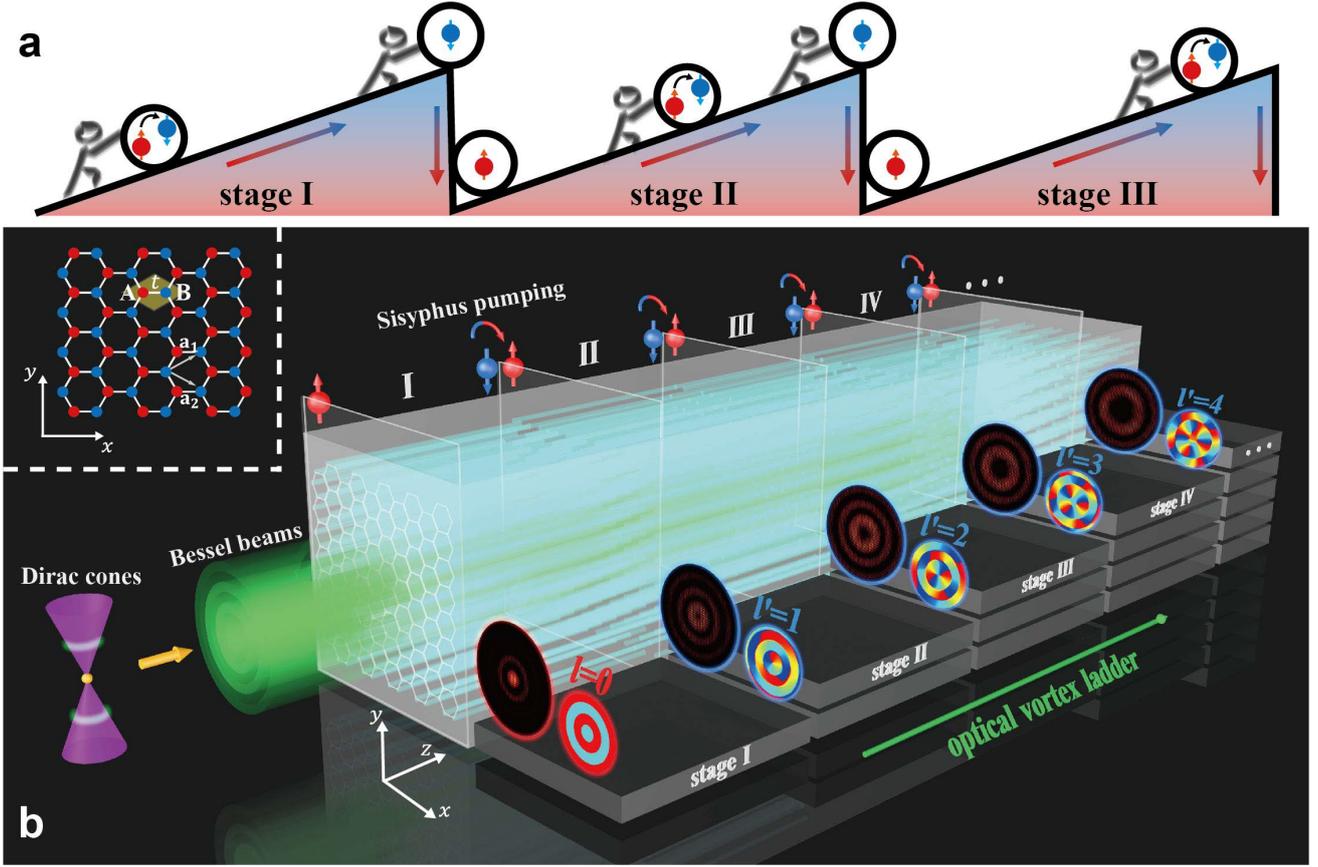

**Fig. 1: Illustration of an optical vortex ladder (OVL) and Sisyphus pumping in photonic graphene.** (a) Schematic of the pseudospin Sisyphus pumping process. The pseudospin modes are represented by the red and blue arrows within the "Sisyphus stone", indicating whether pseudospin-up converts to pseudospin-down (during every stage), or in the opposite direction (between the stages). (b) Excitation of the ring spectra around Dirac cones generates a Bessel beam ($l = 0$) in real space. At stage I of the propagation the beam undergoes a complete conversion from pseudospin-up to pseudospin-down with emerging OAM ($l' = 1$). A judiciously chosen lattice "kick" in between stages I and II implements the Sisyphus pumping of pseudospin-down to pseudospin-up without altering the beam's OAM. Stage II of the propagation is essentially the same as stage I, except that the topological charge after stage II is $l' = 2$. The process repeats itself and forms an OVL that increases OAM by one unit at every stage of the ladder. The structure of photonic graphene is shown in the top-left inset, where the unit cell is highlighted by a shaded rhombus, and $\mathbf{a_1} = \sqrt{3}a/2\,\hat{\mathbf{x}} + a/2\hat{\mathbf{y}}$, $\mathbf{a_2} = \sqrt{3}a/2\,\hat{\mathbf{x}} - a/2\,\hat{\mathbf{y}}$ are the basis vectors with $a$ being the lattice constant.

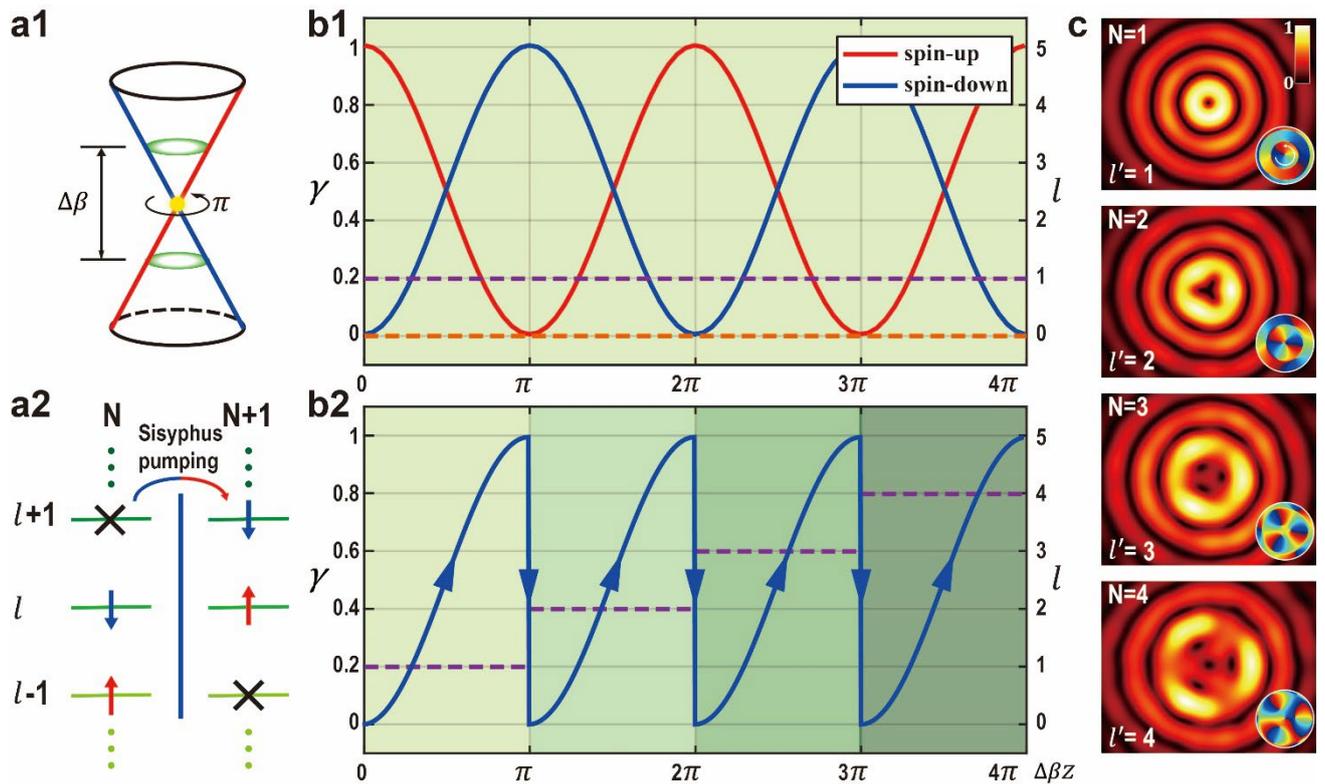

**Fig. 2: Sisyphus pumping with the topological charge climbing up the ladder.** (a1) Excitations of Bloch modes around the Dirac cone with a nontrivial Berry phase in *k*-space, as indicated by the green-shaded region. $\Delta\beta$ represents the well-defined eigenvalue difference. (a2) Illustration of the total topological charges carried in pseudospin modes at adjacent ladder stages ($N$ and $N+1$) due to Sisyphus pumping. The vertical arrows represent the pseudospin mode. (b) The evolution of the pseudospin mode in (b1) the uniform (no Sisyphus pumping) and (b2) the "kicked" (Sisyphus pumping) photonic graphene. The different shaded zones represent the different stages in (b2). The total topological charges carried by the pseudospin down (up) are represented by dashed purple (orange) lines. (c) The associated pseudospin-down components at different stages under Sisyphus pumping, carrying a step increase of topological charges, as depicted by the phase diagram in the insets.

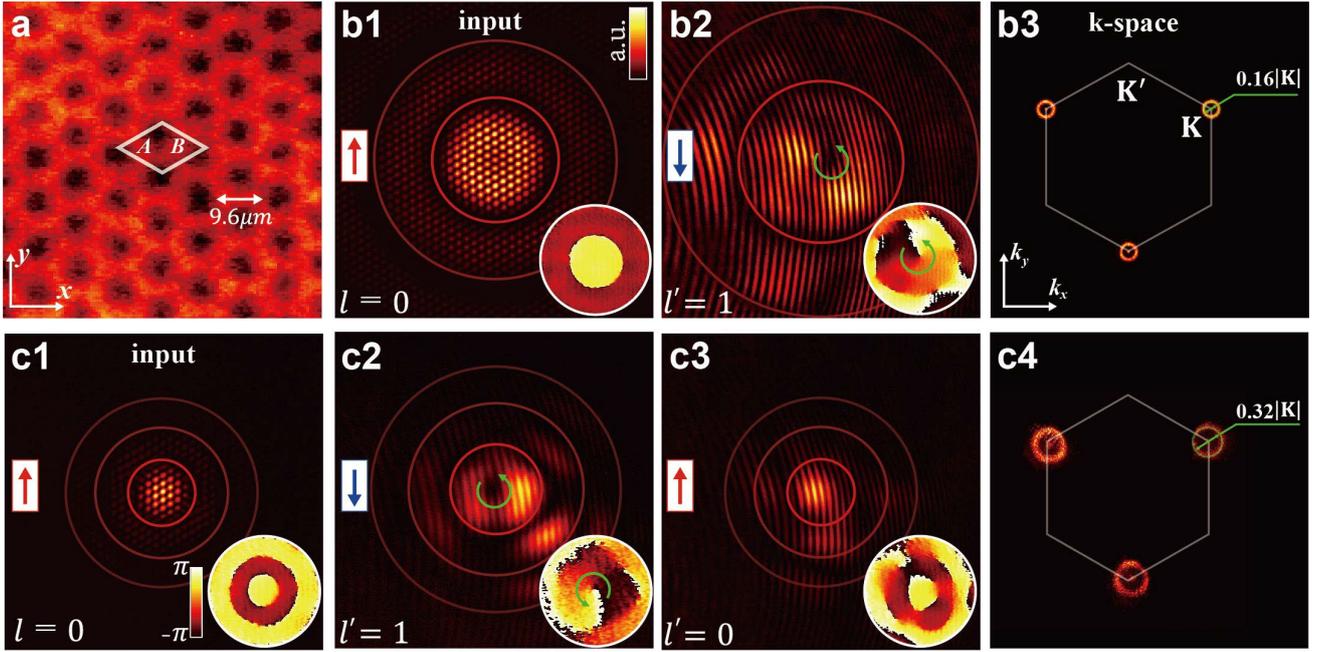

**Fig. 3. Experimental demonstration of a complete pseudospin conversion by spectrum tuning.** (a) An experimentally established photonic graphene lattice with $9.6\mu m$ spacing between the nearest neighbor sites. (b1) Probe pattern of pseudospin-up excitation at three **K** points. The inset is the phase distribution at one of the **K** points. The red circles denote the envelop of the lobes of the Bessel beam. (b2) Output interferogram from one **K** point after 20mm propagation. The position and helicity of the vortex are marked by the curved arrow. (b3) The corresponding *k*-space spectrum with a ring diameter of $d_1 = 0.16|\mathbf{K}|$. (c1) The initial probe beam with a small overall envelope in real space. (c2-c3) Interferograms of output patterns after (c2) 10mm and (c3) 20mm propagation. (c4) The corresponding spectrum with a ring diameter $d_2 = 2d_1$. The vertical arrows represent different pseudospin modes as in Fig. 2.

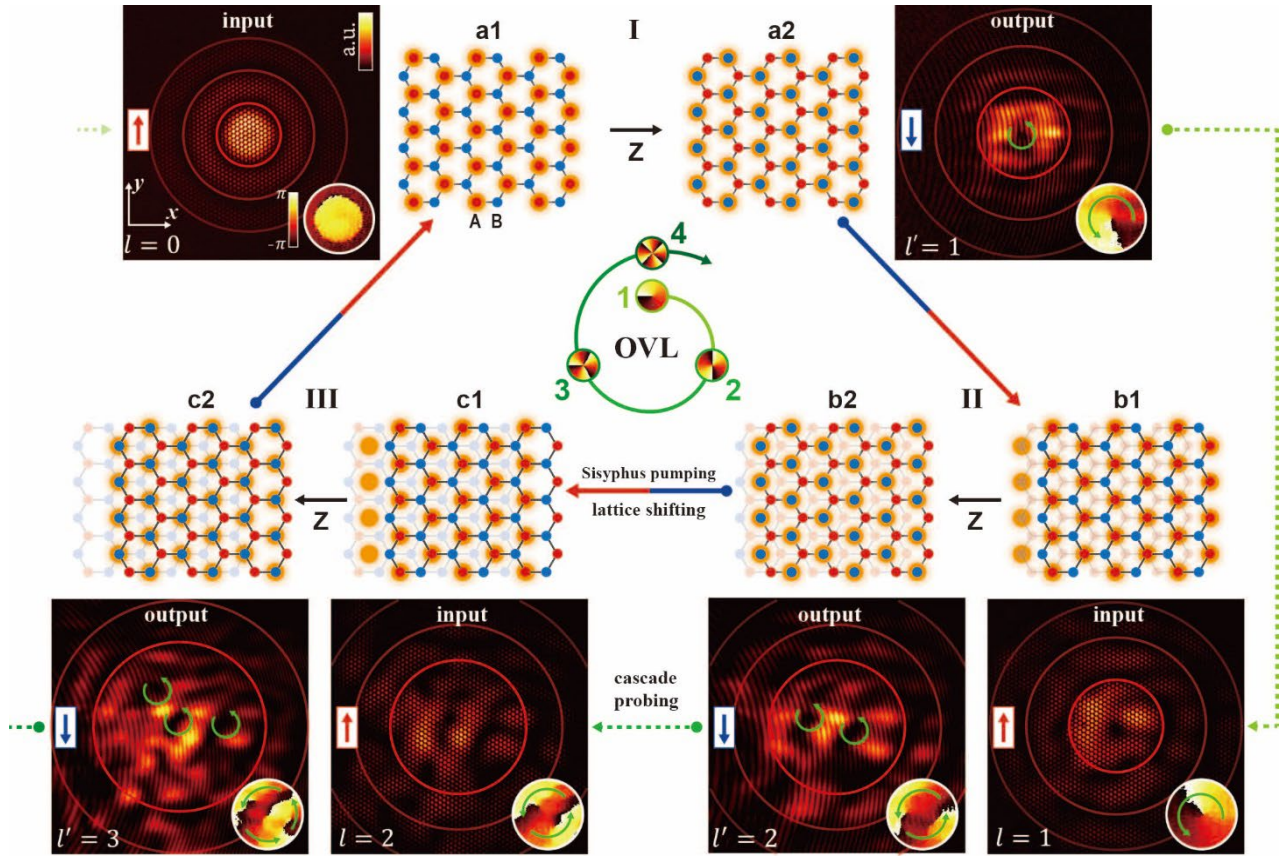

**Fig. 4. Experimental demonstration of the OVL generation via Sisyphus pumping.** The center triangle illustrates the periodic "kicking" of the photonic graphene, while the spiral lines indicate vortices at different OVL stages. The orange dots in the schematic indicate the location of probe beam at the input and output facets. The relative position of the lattices in stage I are shaded for reference. (a) Stage I: The pseudospin-up modes with topological charge ($l = 0$) on sublattice $A$ (a1) transform into pseudo-down modes ($l' = 1$) after a propagation distance $Z = 20mm$, where probe beams transfer entirely to sublattice $B$ (a2). (b) Stage II: The probe beams excite sublattice $A$ through lattice "kicking" and Sisyphus pumping (b1). After propagation, probe beams transfer to sublattice $B$ again (top row in b2), and the topological charge increases to $l' = 2$ (bottom row in b2). (c) Stage III: The probe beams with $l = 2$ located on sublattice $A$ (c1) increase to $l' = 3$ (c2). The inset shows measured phase distribution of the main lobe of the Bessel beam. The position and helicity of the vortex are marked by the curved arrow. The green dashed arrows from input to output beams indicate the cascade probing method.